\documentclass[12pt]{iopart}
\usepackage{graphicx}% Include figure files
\usepackage{amsmath}
\usepackage{amssymb}
\usepackage{color}
\usepackage{empheq}
\usepackage{xcolor}
\usepackage{cancel}
\usepackage{siunitx}
\usepackage{tcolorbox}
\usepackage{url}

\begin{document}

\title[QSHS: An Axion Dark Matter Resonant Search Apparatus]{QSHS: An Axion Dark Matter Resonant Search Apparatus}

\author{
A.\,Alsulami\textsuperscript{a}, 
I.\,Bailey\textsuperscript{b}, 
G. Carosi\textsuperscript{c}, 
G.\,Chapman\textsuperscript{d}, 
B.\,Chakraborty\textsuperscript{e},
E.\,J.\,Daw\textsuperscript{f}\footnote{\texttt{e.daw@sheffield.ac.uk}},
N.\,Du\textsuperscript{c},
S.\,Durham\textsuperscript{c},
J.\,Esmenda\textsuperscript{b}
J.\,Gallop\textsuperscript{d},
T.\,Gamble\textsuperscript{f},
T.\,Godfrey\textsuperscript{e}
G.\,Gregori\textsuperscript{a}
J.\,Halliday\textsuperscript{a}
L.\,Hao\textsuperscript{d},
E.\,Hardy\textsuperscript{a,g}
E.A.\,Laird\textsuperscript{b},
P.\,Leek\textsuperscript{a},
J.\,March-Russell\textsuperscript{a},
P.\,J.\,Meeson\textsuperscript{h},
C.\,F.\,Mostyn\textsuperscript{f},
Yu.\,A.\,Pashkin\textsuperscript{b},
S.\,Ó.\,Peatain\textsuperscript{b},
M.\,Perry\textsuperscript{f},
M.\,Piscitelli\textsuperscript{a},
M.\,Reig\textsuperscript{a,h}
E.\,J.\,Romans\textsuperscript{e,j},
S.\,Sarkar\textsuperscript{a},
P.\,J.\,Smith\textsuperscript{f},
A.\,Sokolov\textsuperscript{a},
N.\,Song\textsuperscript{i},
A.\,Sundararajan\textsuperscript{f},
B.-K\,Tan\textsuperscript{a},
S.\,M.\,West\textsuperscript{h},
S.\,Withington\textsuperscript{a}
}
\address{\textsuperscript{a}Department of Physics, University of Oxford, Parks Road, Oxford OX1 3PU, UK}
\address{\textsuperscript{b}Department of Physics, Lancaster University, Lancaster LA1 4YB, UK}
\address{\textsuperscript{c}Lawrence Livermore National Laboratory, Livermore, California 94550, USA}
\address{\textsuperscript{d}National Physical Laboratory, Hampton Road, Teddington TW11 0LW, UK}
\address{\textsuperscript{e}The London Centre for Nanotechnology, 17-19 Gordon Street, London WC1H 0AH, UK}
\address{\textsuperscript{f}School of Mathematical and Physical Sciences, The University of Sheffield, Hicks Building, Hounsfield Road, Sheffield S3 7RH, UK}
\address{\textsuperscript{g}Department of Mathematical Sciences, University of Liverpool, Liverpool, L69 7ZL, UK}
\address{\textsuperscript{h}Royal Holloway University of London, Egham, Surrey, TW20 0EX, UK}
\address{\textsuperscript{i}Institute of Theoretical Physics, Chinese Academy of Sciences, Beijing, 100190, China}
\address{\textsuperscript{j}Department of Electronic and Electrical Engineering, University College London, London WC1E 6BT, UK}

\vspace{10pt}
\begin{indented}
\item[]April 16\textsuperscript{th} 2025
\end{indented}

\begin{abstract}
We describe a resonant cavity search apparatus for axion dark matter constructed by the 
Quantum Sensors for the Hidden Sector (QSHS) collaboration. The apparatus is configured to search for 
QCD axion dark matter, though also has the capability to detect axion-like particles (ALPs), dark photons,
and some other forms of wave-like dark matter. Initially, a tuneable
cylindrical oxygen-free copper cavity is read out using a low noise microwave amplifier feeding
a heterodyne receiver. The cavity is housed in a dilution refrigerator and threaded
by a solenoidal magnetic field, nominally $\rm 8\,T$. The apparatus also houses a magnetic field 
shield for housing superconducting electronics, and several other fixed-frequency resonators for
use in testing and commissioning various prototype quantum electronic devices sensitive at a range
of axion masses in the range $\rm 2.0$ to $\rm 40\,{\mathrm \mu} eV/c^2$. The apparatus as currently
configured is intended as a test stand for electronics over the relatively wide frequency band attainable
with the $\rm TM_{010}$ cavity mode used for axion searches. We present performance data for the 
resonator, dilution refrigerator, and magnet, and plans for the first science run.
\end{abstract}

%\maketitle

\section{Physics Motivation}
\label{sec:introduction}

Almost a century after Zwicky's observations of excess galactic virial velocities in the Coma Cluster \cite{1933AcHPh...6..110Z}, the dark matter problem remains one of the most critical unsolved mysteries of fundamental physics \cite{RevModPhys.90.045002}. Presently, the existence of dark matter is inferred from multiple observations involving its gravitational interactions, most significantly its effects on the rotation of spiral galaxies \cite{Yoon_2021}, lensing of background light from clusters \cite{Wambsganss:1998gg}, and modeling of the cosmic background radiation \cite{ParticleDataGroup:2020ssz}.  Direct, non-gravitational detection of dark matter
is challenging because of its feeble
interactions with everyday matter and because its precise nature is yet undetermined.  

A leading candidate is a class of particles known as \emph{axions}.
Axions originally arose \cite{Weinberg:1977ma,Wilczek:1977pj} as a consequence of Peccei and Quinn's postulated solution~\cite{Peccei:1977hh} to the strong CP problem of QCD. It has now been realised that high-scale theories, such as string theory, imply that this original QCD axion can be accompanied by many variant axions with broadly similar couplings but with a wide range of possible masses~\cite{Svrcek:2006yi,Conlon:2006tq,Arvanitaki:2009fg}.
Moreover, the details of the relationships between the axion couplings to normal Standard Model matter and the axion mass 
provide a uniquely sensitive probe of the physics of the smallest distance (equivalently largest energy) scales, approaching the GUT or even Planck scale~\cite{Kamionkowski:1992mf,Barr:1992qq,Holman:1992us,Agrawal:2022lsp,Agrawal:2024ejr}. Importantly, axions may naturally be the dominant, or a significant component of, dark matter~\cite{Abbott:1982af,Preskill:1982cy,Dine:1982ah}. 

Among the possible QCD axion (and ALP) 
couplings, probably the most useful
for searches is the axion-photon coupling defined by
\begin{align}
\Delta\mathcal{L}_{\rm int}=\frac{g_{\gamma}\alpha }{\pi f_{a}}\varepsilon_0a(x) {\bf E}(x)\cdot{\bf B}(x) ~,
\label{eq:axion-photon-coupling}
\end{align}
where $a(x)$ and ${{\bf E}(x)}, {{\bf B}(x)}$ are the axion and EM fields, $\alpha$
is the fine structure constant, and $f_a$ is the unknown ``axion decay constant", a dimensionful parameter of the theory whose inverse sets the overall strength of all axion interactions.  The dimensionless parameter $g_\gamma$ is a model-dependent factor setting the precise axion-photon coupling including the effect of mixing with other pseudoscalar states such as the $\pi^0$ meson.  In almost all models $g_\gamma \sim {\cal O}(1)$, for example, the DFSZ model \cite{Dine:1981rt,Zhitnitsky:1980tq} yields $g_\gamma \approx -0.37$, while the so-called KSVZ ``hadronic" model \cite{Kim:1979if,Shifman:1979if} gives $0.97$ (however, a variant model of KSVZ has a suppressed $g_\gamma \approx -0.04(2)$ with the last digit uncertainty indicated). At the macroscopic level, the effect of the coupling Eq.(\ref{eq:axion-photon-coupling}) in free space is to modify two of the Maxwell equations by the the addition of new axion-dependent effective charge and current densities, $\rho_a$ and ${\bf j}_a$ to the conventional electronic charge and current densities,
\begin{align}
    \nabla \cdot (\varepsilon_0 {\bf E}) = \rho_e + \rho_a, \\
    \nabla \times ({\bf B}/\mu_0) - \partial_t (\varepsilon_0 {\bf E}) = {\bf j}_e + {\bf j}_a,
    \label{eq:axion_electrodynamics}
\end{align}
where the background electric and magnetic field dependent expressions for $\rho_a$ and ${\bf j}_a$ are
\begin{align}
    \rho_a &\equiv g_{a\gamma\gamma} \sqrt{\frac{\varepsilon_0}{\mu_0}}{\bf B}\cdot\nabla a, \\
    {\bf j}_a &\equiv g_{a\gamma\gamma}\sqrt{\frac{\varepsilon_0}{\mu_0}}({\bf E}\times \nabla a - {\bf B} \frac{\partial a}{\partial t}) ~.
    \label{eq:axion_sources}
\end{align}
Here the effective axion coupling to EM that determines the signal in the apparatus is
\begin{align}
    g_{a\gamma\gamma} \equiv \frac{g_\gamma \alpha}{\pi f_a}~.
    \label{eq:axion-photon-effective-coupling}
\end{align}
This modification of Maxwell theory in the presence of an axion field is known as ``axion electrodynamics"~\cite{Tobar:2018arx}.

For the QCD axion the value of $f_a$ sets the mass, $m_a$, of the axion, with latest precision calculations giving~\cite{GrillidiCortona:2015jxo}
\begin{align}
m_a c^2 = 5.70(7)\,\, {\rm \mu eV} \,\left(\frac{10^{12}\, {\rm GeV}}{f_a}\right) ~.
\label{eq:QCDaxionmass}
\end{align}

Soon after the invention of the original Peccei-Quinn-Weinberg-Wilzek axion models it was realized that collider and precision constraints force $f_a$ to be a superheavy energy scale, with, now, the analysis of observations of hot astrophysical objects giving the bound $f_a\gtrsim 10^8~{\rm GeV}$~\cite{ParticleDataGroup:2020ssz}.
This implies that axions are very feebly coupled to each other and to all Standard Model states, and are thus both extremely long-lived if they are low-mass and ``dark" (i.e., not substantially interacting with photons) and so a good dark matter candidate.

In the regime of axion masses of greatest
interest to us, the mass of the axion particles is so tiny ($\sim 10^{-5}{\rm eV/c^2}$) that the inferred number density of dark matter particles in our galaxy is so large that the axion dark matter is better thought of as a \emph{wave-like field} with coherence length and coherence time set by the galactic dynamics. 

Utilising the axion-photon coupling, Eq.(\ref{eq:axion-photon-coupling}),
the most sensitive direct detection experiments employ the method of Sikivie \cite{Sikivie:1983ip,Sikivie:2020zpn}, wherein halo axions convert to photons in an electromagnetic resonator threaded by a static magnetic field. Assuming the relation between $f_a$ and $m_a$ given in Equation (\ref{eq:QCDaxionmass}), the projected signal power in such apparatus is given by 
\begin{align}
    P_{a\rightarrow\gamma}&=1.79\times10^{-21}\,{\rm W}\left(\frac{V}{\rm 220\,L}\right)
    \left(\frac{B}{\rm 7.6\,T}\right)^2
    f_{\rm nlm}\left(\frac{g_\gamma}{0.97}\right)^2 \nonumber \\
    &\times \left(\frac{\rho_a}{\rm 0.45\,GeV/cc}\right)\left(\frac{\nu_a}{\rm 750\,MHz}\right)
    \left(\frac{Q}{70,000}\right),
    \label{eq:axsigpower}
\end{align}
where $V$ is the cavity volume, $B$ is the applied magnetic field, $\rho_a$ is the local energy density of halo dark matter, $\nu_a$ is the frequency of the photons produced and $Q$ is the unloaded quality factor of the resonant mode, under the assumption that the $Q$ factor of the axions themselves is significantly higher. The quantity $f_{\rm nlm}$ is a form factor defined by
\begin{align}
f_{\rm nlm}=\frac{\left(\int {\bf E}_{\rm nlm}(x)\cdot {\bf B}(x)\,dV\right)^2}{\left(\int |{\bf E_{\rm nlm}}|^2\,dV\right)\left(\int |{\bf B}|^2\,dV\right)},
\end{align}
where the integrals are over the cavity volume, ${\bf B}(x)$ is the applied magnetic field and ${\bf E}_{\rm nlm}(x)$ is the electric field of the cavity mode \cite{Jackson1998}. The axion linewidth is at most that of a distribution thermalised in the gravitational potential of our galaxy, with r.m.s velocity $v_0$ of around $\rm 230\,km\,s^{-1}$, leading to an axion $Q$-factor of order $10^7$. This is significantly greater than quality factors achievable in normal-conducting cavities used in the high magnetic fields employed by cavity axion haloscopes, though it is possible that field tolerant superconducting coatings may enable significantly higher Qs in the future.

In the QSHS initial prototype, where the focus is on a high bandwidth so that we can test technologies rather than initially on high sensitivity, we have a much smaller cavity, of volume $\rm 0.556\,litres$, containing a larger tuning rod, leading to a smaller $\rm TM_{010}$ mode quality factor. Written with dimensionful quantities reflecting the current QSHS parameters, the projected signal power is 
\begin{align}
    P_{a\rightarrow\gamma}&=1.43\times 10^{-24}\,{\rm W} \left(\frac{V}{\rm 0.556\,{\rm L}}\right)
    \left(\frac{B}{\rm 8\,T}\right)^2
    f_{\rm nlm}\left(\frac{g_\gamma}{0.97}\right)^2 \nonumber \\
    & \times \left(\frac{\rho_a}{\rm 0.45\,GeV/cc}\right)\left(\frac{\nu_a}{\rm 5\,GHz}\right)
    \left(\frac{Q}{3,000}\right).
    \label{eq:axsigpowerqshs}
\end{align}

The small signal power challenges experimentalists aiming to detect photons from axion conversion in the cavity. The signal bandwidth is \mbox{$\Delta\nu\sim\nu_a(v_0^2/2c^2)$}. For a photon frequency of $\rm 5\,GHz$, \mbox{$\Delta \nu\sim{\rm 1.6\,kHz}$}. The temperature $T_N$ of a Johnson noise source equivalent to this signal is given by $k_B T_N\Delta\nu=P_S$, where $P_S$ is the signal power in the receiver electronics, $P_S=P_{a\rightarrow\gamma}/4$. The factor of $1/4$ results from degrading the quality factor of the cavity by a factor of $2$ when it is critically coupled to the receiver electronics, and only half of this power is deposited in the receiver. Assuming as an example the signal power given as a prefactor in Equation (\ref{eq:axsigpower}), $T_N\sim{\rm 20\,mK}$, and for the signal power in Equation (\ref{eq:axsigpowerqshs}) as $\rm 16\,\mu K$. Such experiments therefore run at cryogenic temperatures, and utilise state of the art quantum measurement techniques to minimise the noise contributions from the physical temperature $T_C$ of the cavity walls, and the noise temperature $T_A$ of the receiver electronics, respectively.

The signal is a bandwidth peak significantly narrower than the structure in the power spectrum of the receiver output resulting from the cavity resonance. Potential axion signals can be distinguished from most sources of background by requiring that the signal strength is proportional to $B^2$. The signal to noise ratio, $\rm SNR$, is given by the ratio of the signal power to the statistical fluctuations in the noise between neighbouring bins in the power spectrum at the receiver output, and can be estimated from the radiometer equation,
\begin{align}
    {\rm SNR}=\frac{P_S}{k_B(T_C+T_A)}\sqrt{\frac{\Delta t_I}{\Delta f}},
    \label{eq:radiometer}
\end{align}
where $\Delta f$ is the resolution bandwidth of each bin in the power spectrum of the receiver output and $\Delta t_I$ is the integration time. $\Delta f$ is typically narrower than the signal bandwidth $\Delta \nu$ to avoid losing sensitivity for signals that straddle adjacent frequency bins. The resonant cavity must be tuned to overlap each frequency at which the search is carried out, and $t_I$ must be sufficiently long to achieve $\rm SNR$ of order 4 to have sufficient confidence of detecting a signal.  

\section{Overview of the Apparatus}
\label{sec:overview}

\begin{figure}[h!]
    \centering
    \includegraphics[width=8.3cm]{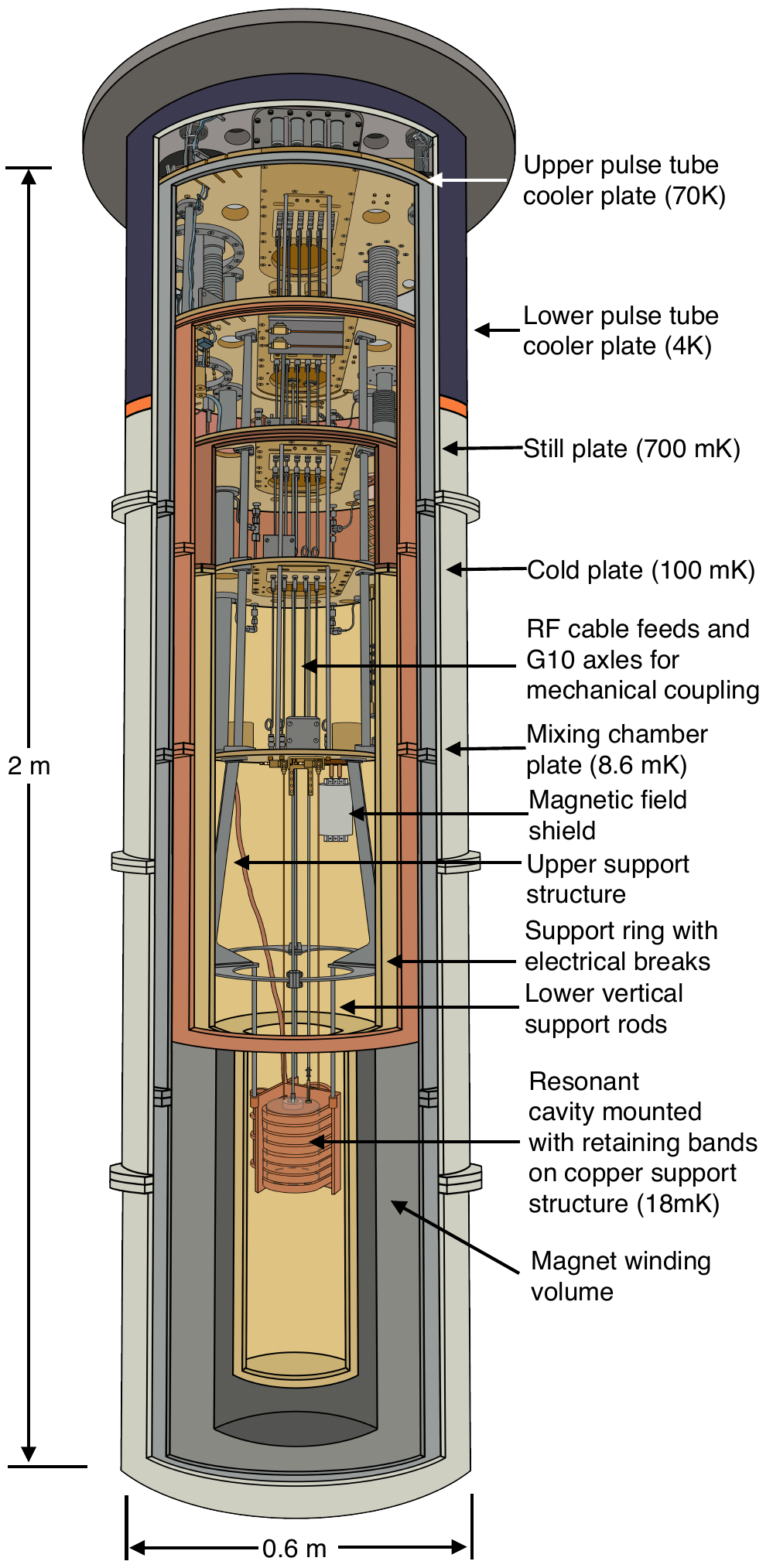}
    \caption{A schematic of the QSHS cryostat and detector. The dilution refrigerator is an Oxford Instruments Proteox MX utilising mechanical two stage pulse-tube precooling and a $\rm ^3He/^4He$ dilution refrigerator. The magnet is an $\rm 8\,T$ three-coil superconducting solenoid. }
    \label{fig:firstschematic}
\end{figure}

The QSHS facility has been assembled in a newly refurbished laboratory at the University of Sheffield. The facility houses a Proteox MX dilution refrigerator (DF) and a superconducting solenoidal magnet, supplied as an integrated system by Oxford Instruments. Figure \ref{fig:firstschematic} is a schematic of the apparatus.

The DF mixing chamber plate reaches a base temperature of $\rm 8.6\,mK$. The magnet has a maximum field strength of $\rm 8\,T$ and encloses a cylindrical payload space $\rm 18.5\,cm$ in diameter and $\rm 20\,cm$ deep. Above the magnet bore, there is a second volume of diameter $\rm 40\,cm$ and height $\rm 37\,cm$. Due to the outer shield coil of the magnet, the field in this upper payload region falls off to below $\rm 10^{-2}\,T$ within $\rm 16\,cm$ of the mixing chamber plate, allowing the installation of passive magnetic field shields in this space to house ultra-low-noise electronics that requires a very low magnetic field environment. The first of these field shields is described in Section \ref{sec:fieldshield}.

A moveable clean room to mitigate dust contamination forms a working area around the dilution refrigerator when dismantled. The cleanroom incorporates a fan providing filtered positive-pressure airflow through a HEPA filter. 

The axion target consists of a cylindrical resonant space hollowed out from oxygen-free copper to form a cavity. The cavity contains a hollow copper cylindrical tuning rod mounted parallel to the cavity axis on off-centre axles passing through holes in the top and bottom of the cavity, captured by ceramic bearings. Rotation of these axles moves the tuning rod in a circular arc so that its distance from the central axis changes, altering the frequencies of the cavity $\rm TM$ modes. A $\rm TM_{010}$ mode tuning range of $\rm 4.1-7.5\,GHz$ can be achieved. The initial cavity and tuning rod geometry is optimised for testing a wide variety of superconducting electronics for use in axion searches in the mass range $\rm 17-31\,\mu eV/c^2$. The cavity hangs from the mixing chamber plate by a stainless steel support frame. A soft copper thermal link provides a path for cooling of the cavity by the dilution refrigerator.

Electromagnetic coupling to the cavity modes is accomplished using a semi-rigid coaxial electric field probe whose outer conductor is in electrical contact with a beryllium copper spring fingerstock mounted in a threaded hole on the top of the cavity. The inner conductor protrudes several millimetres into the cavity space. Adjustment of the insertion depth of the field probe and the tuning rod position are discussed in Section \ref{sec:CavityMechanisms}.

The data acquisition system (DAQ) consists of slow controls and monitoring for temperatures, pressures and voltages associated with experiment operations, and a fast data acquisition system for data from the cavity and RF electronics. The DAQ system also communicates with the controls for the dilution fridge and magnet. Relevant slow controls information is incorporated as a header in data acquired from the electronics reading out the cavity. Cavity readout electronics consists of an ultra-low-noise heterodyne receiver, further room temperature amplifiers, a bandpass filter, and a $\rm 125\,MHz$ digitiser. The digitised data is decimated and further heterodyned to an audio band matched to the bandwidth of the cavity $\rm TM_{010}$ mode. Welch estimates of the power in each data segment and slow controls readout constitute the raw data.

\section{Cryostat}
\label{sec:cryostat} 

QSHS has purchased a Proteox MX500 system from Oxford Instruments \cite{oxinstproteox}. This dry closed-cycle refrigerator is housed in an aluminium outer vacuum envelope whose cylindrical component sections are removed from below. Vacuum seals are achieved with Viton `O' rings. The dry design means that only the outer room-temperature envelope and the inner closed-loop helium systems are required to be vacuum tight. A Pfeiffer ASM-340 helium leak checker is used to verify the integrity of the outer vacuum envelope after pumping down and before cooling is initiated. The upper (PT1) and lower (PT2) pulse tube cooler plates, shown in Figure \ref{fig:firstschematic} reach base temperatures of $\rm 77\,K$ and $\rm 4\,K$, respectively. Pre-cooling is achieved with a Sumitomo RP-182B2S two-stage pulse tube cooler having $\rm 1.5\,W$ of cooling power at a PT2 temperature of $\rm 4\,K$, driven from a 3-phase Sumitomo Heavy Industries F-100 compressor \cite{sumitomo}. 

The dilution refrigerator requires an 18 litre charge of $\rm ^3He$ in a $\rm 100$ litre vessel of $\rm ^3He/^4He$ gas mixture supplied by Oxford Instruments. Figure \ref{fig:firstschematic} shows the positions of the still, cold, and mixing chamber plates. The cooling power of the fridge at the mixing chamber plate is $\rm 450\,\mu W$ at $\rm 100\,mK$ and $\rm 12\,\mu W$ at $\rm 20\,mK$. A Lakeshore 372 AC resistance bridge and Lakeshore ruthenium oxide temperature sensors allow for cryogenic temperature control \cite{lakeshore}.

Thermal shields are attached to upper pulse tube (PT1), lower pulse tube (PT2), and still plate stages, minimising radiative heat transfer to the experimental volume. The fridge incorporates a solenoidal magnet, also supplied by Oxford Instruments and integrated into the cryostat. This magnet will be described in detail in Section \ref{sec:magnet_design}. The magnet is suspended from the bottom of the copper $\rm 4\,K$ thermal shield attached to PT2. The inner bore of the magnet is $\rm 200\,mm$ with a $\rm 185\,mm$ diameter working space inside the thermal baffle between the lower payload space and the inner wall of the magnet. The magnet has a physical bore length of $\rm 428\,mm$ and a usable bore length of $\rm 300\,mm$, between the top of the magnet bore and the cylindrical bottom plate of the $\rm 700\,mK$ shield. At the wider upper payload space, referred to in Section \ref{sec:overview}, the thermal shield has an inner diameter of $\rm 360\,mm$. 

Electrical connections into the fridge consist of three low frequency looms, one of these looms being split into two different wiring configurations and intended for piezo drives that may in future be deployed on the apparatus. There are sixteen RG405 semi-rigid coaxial lines. The electrical connections are summarised in Table \ref{tab:cables}. In addition, there are electrical looms used exclusively for dilution fridge controls which will not be described here. For the latter, the reader is referred to Oxford Instruments \cite{oxinstproteox}.

\begin{table}[h!]
    \centering
    \begin{tabular}{|p{25mm}|p{15mm}|p{20mm}|p{20mm}|p{20mm}|}
    \hline
    Cable Set & BW & RT--PT2 & PT2-- Mix. Ch. & Conductors \\
    \hline \hline
    1. Low BW twisted pairs & DC to $\rm 10\,kHz$ 
    & constantan & NbTi & 12 twisted pairs \\ \hline
    2. Low BW twisted pairs & DC to $\rm 10\,kHz$
    & constantan & NbTi & 12 twisted pairs \\  \hline
    3. Low BW twisted pairs & DC to $\rm 10\,kHz$ 
    & constantan & constantan & 8 twisted pairs \\ \hline
    4. Low BW twisted pairs & DC to $\rm 10\,kHz$
    & copper & phosphor-bronze, $\parallel$ NbTi 
    & 4 twisted pairs \\ \hline
    5. Unattenuated RG405 & DC to $\rm 18\,GHz$
    & beryllium copper & NbTi & 4 coaxial \\ \hline
    6. Attenuated RG405 & DC to $\rm 18\,GHz$
    & stainless steel & stainless steel & 12 coaxial \\ \hline
    \end{tabular}
    \caption{Electrical cables in the dilution refrigerator. Coaxial cables are RG405 terminated with SMA connectors throughout. Feedthroughs at each flange serve as thermal
    couplings. The attenuated lines have in-line $\rm 20\,dB$ attenuators at PT2 and cold plates and $\rm 10\,dB$ at the mixing chamber plate. The low bandwidth twisted pairs feed through the room temperature flange on 24 way Fischer connectors, and are thermally sunk at each plate using gold plated copper clamps around the ribbons.}
    \label{tab:cables}
\end{table}

For extra capacity, there are 12 unused SMA feedthroughs on the room temperature flange, as well as 3 unused Fischer connector feedthroughs for additional low frequency wiring. A fourth Fischer connector is used for the wires to the piezo motor drives described in Section \ref{sec:CavityMechanisms}. All sixteen RF coaxial lines and the four extra-capacity Fischer connector feedthroughs are mounted on a swappable rectangular cartridge that interfaces with all the plates in the system. A large portion of the electrical wiring and experimental configuration may be swapped out with other configurations for efficient use of the DF by more than one experimental group.

\begin{figure}
    \centering
    \includegraphics[width=8.3cm]{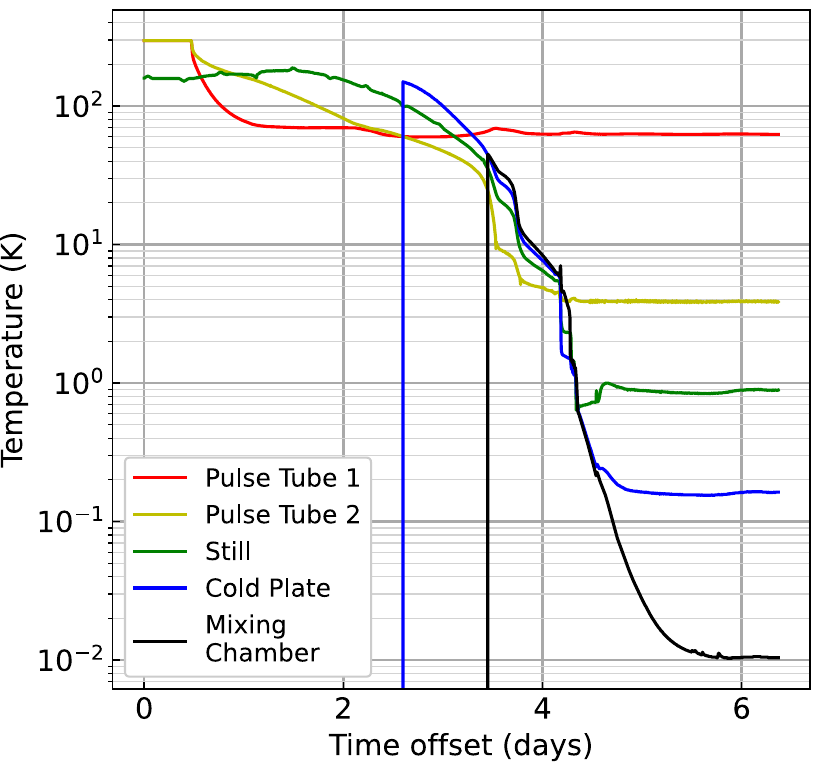}
    \caption{Cooling curves for the dilution refrigerator with the cavity and frame mounted, but without the tuning rod installed. The cavity reached $\rm 18.6\,mK$. Details of the temperature sensors and readout are given in Section \ref{sec:cryostat}. The cold plate and mixing chamber sensors read zero until time offsets of 2.7 and 3.3 days respectively.}
    \label{fig:cooling_curve}
\end{figure}

Figure \ref{fig:cooling_curve} shows temperatures at the two pulse tube and three dilution refrigerator stages during the first cooling run where the cavity and frame discussed in Sections \ref{sec:frame} and \ref{sec:cavity} were attached to the mixing chamber plate. The empty cavity reached an operating temperature of $\rm 18.5\,mK$. Future runs will determine the temperature reached when the tuning rod is installed.

\section{Support Frame}
\label{sec:frame}

\begin{figure}[htbp]
    \centering
    \includegraphics[width=8.3cm]{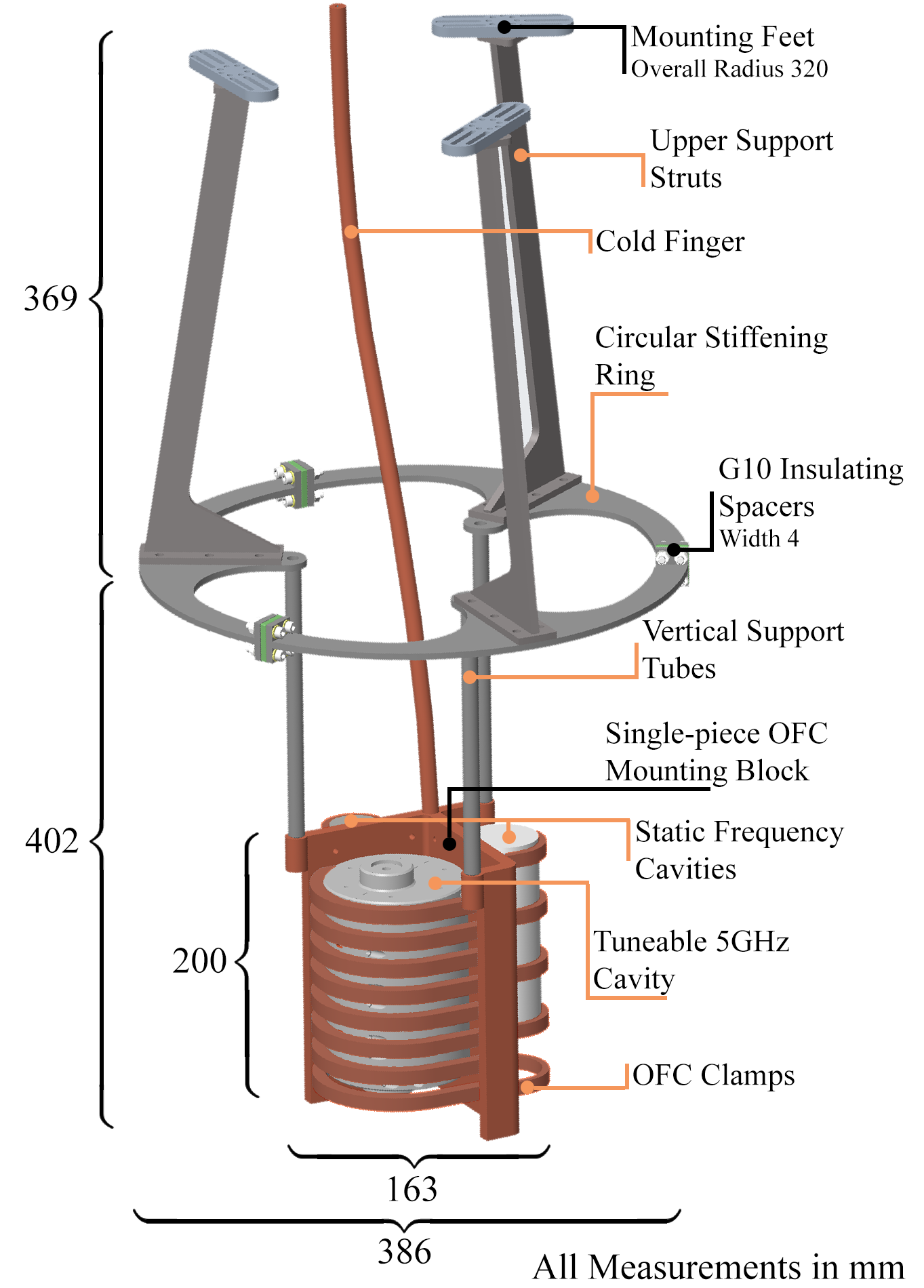}
    \caption{Rendered drawing of the cavity support frame. Dimensions are in $\rm mm$.}
    \label{fig:Support-Frame}
\end{figure}

The cavity is mounted with its symmetry axis vertical, with its centre $\rm 677\pm5\,mm$ below the mixing chamber plate. The support frame is fabricated from grade 316 stainless steel, and utilises a three--leg geometry for rigidity and ease of access. A 3-dimensional rendering of the support structure is shown in Figure \ref{fig:Support-Frame}. The geometry consists of three rectangular cross section upper support struts that hold a circular stiffening ring, occupying almost the full diameter of the $\rm 400\pm2\,mm$ diameter upper payload volume. This ring incorporates three support points defining a smaller diameter, $\rm 195\pm1\,mm$, for three vertical cylindrical support tubes that descend into the magnet bore. Mounting to the mixing chamber plate is via the existing matrix of M3 threaded mounting holes, with stainless steel nuts and washers on the upper side of the mixing chamber plate providing additional vertical support and guard against stripping of the copper M3 threaded mounting holes.

The vertical support tubes are bolted at their bottom ends to a single-piece oxygen-free copper vertical support frame to which the cavity payload is clamped. The same grade of copper is used for the bands, having threaded holes at either end. These bands are bolted to the frame and clamp the oxygen-free resonant cavity in its operating position. Mounting holes are also provided for other cavities having higher resonant frequencies to be attached to the support frame as shown in Figure \ref{fig:Support-Frame}. A soft copper rod of circular cross section is securely bolted to the mixing chamber plate, and bolted with a conical-geometry mating face to the copper support frame. This arrangement provides a high thermal conductivity path between the mixing chamber and the cryogenic payload. The copper rod is bent to allow for differential thermal expansion between the steel frame and the copper rod. 

Magnetic field quench protection is through the rigid and mechanically strong stainless steel support structure, with additional precautions including G10 insulating breaks in the circular supporting ring and predominantly vertical orientation of the support structures. Several threaded M3 holes in the copper cavity and support frame allow thermometers to be attached.

\section{Magnet}
\label{sec:magnet_design}

\begin{figure}[htbp]
\centering
\includegraphics[width=8.3cm]{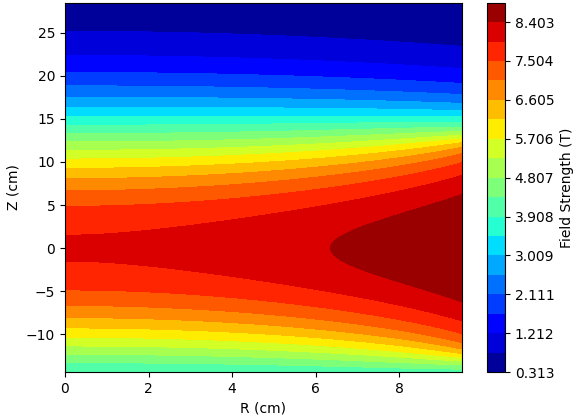}
\caption{Magnetic field in and directly above the bore, where $(R,Z)=(0,0)$ is the centre. The data was provided by Oxford Instruments.\cite{oxinstproteox}}
\label{fig:Bore_Field}
\end{figure}

The magnet was supplied integrated into the cryostat by Oxford Instruments. It is conductively cooled via the copper thermal shield connecting the outer diameter of its top flange to the PT2 plate. The magnet is a dry superconducting design. The inner coil is niobium-tin, and the outer two coils are niobium-titanium. Current is supplied to the three coils in series. A heat switch permits persistent operation. External power supplies provided by Oxford intruments provide the $\rm 160\,A$ of current necessary to reach full field, which is $\rm 8\,T$ at the centre of the bore. The field is nominally cylindrically symmetric, although outside the magnet bore the off-centre quantum electronics field shield breaks the cylindrical symmetry in its vicinity. Figure \ref{fig:Bore_Field} is a simulated field map inside the magnet provided by Oxford Instruments. The fields of superconducting magnets in persistent mode are very time-stable, and we do not anticipate contributions to the noise from fluctuations in the magnetic field itself. Care must however be taken to prevent wires from moving in the applied field. In particular, the low frequency cables connected to temperature sensors can pick up noise due to induced currents. Such cables are therefore tied firmly to supports and fixtures to minimise such vibrations. 

The stray field is kept minimal by the outer niobium-titanium windings, which act as a magnetic field shield. The stray field  falls to $\rm 5\,G$ at $\rm 1.2\,m$ and $\rm 1.4\,m$, respectively, in the horizontal plane and along the axis from the centre of the magnet bore. The well confined high field region ensures that stray field does not affect neighbouring labs and corridors, or couple to the rebar in the concrete floor. It also provides a low field environment for workers and for electronics connected the fridge, allowing cable lengths to be minimal, and permitting operators to work without concerns about field over the majority of the lab and at all the control racks. Short cables are advantageous for minimising microwave signal losses and for preserving the sawtooth waveforms used to drive the piezo motors described in Section \ref{sec:CavityMechanisms}.

The ramp rate of the magnet loaded with the cavity and frame is typically, $\rm 0.2\,A/min.$ or $\rm 0.01\,T/min$. This  rate is arrived at as a compromise between the requirements of minimising the total ramp time to full field, and preventing the temperature of the payload exceeding $\rm 100\,mK$ due to eddy current heating.

\section{Resonators}
\label{sec:resonators}

\subsection{The QSHS Cavity}
\label{sec:cavity}

\begin{figure}[htbp]
    \centering
    \includegraphics[width=8.3cm]{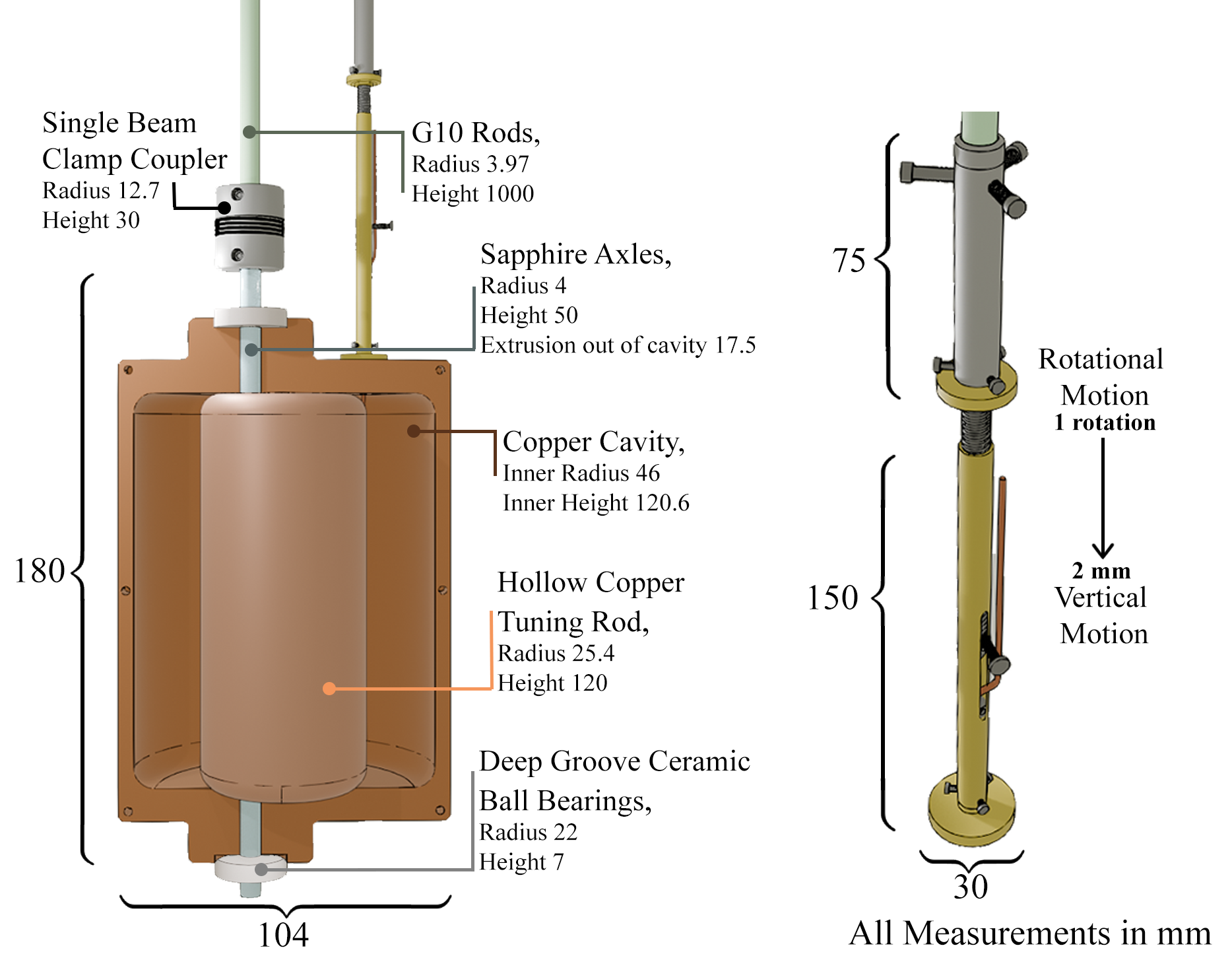}
    \caption{Diagram of one half of the copper cavity and the inner tuning and coupling mechanisms (left), a zoomed view of the rotational-to-linear antenna coupling mechanism is shown on the right.}
    \label{fig:CavitySchematic}
\end{figure}

For QSHS's initial run, a design based on that of the ADMX sidecar cavity is used. The QSHS-adapted version of this, including the overall cavity dimensions and tuning mechanisms are shown in Figure \ref{fig:CavitySchematic}. The cavity was clamped to the support plate described in Section \ref{sec:frame} with its major axis vertical. Rotating the tuning rod off-centre from the cavity tunes the transvere magnetic (TM) mode frequencies. In the absence of a tuning rod, most axion haloscopes use the most fundamental of these, the $\rm TM_{010}$ mode \cite{Jackson1998}, as it results in the largest overlap between the cavity mode’s electric field and the external magnetic field, and hence the largest
axion signal strength. In the presence of a tuning rod the modes of the cavity hybridize and no longer correspond precisely to the modes of an empty cavity, but in this paper we follow the usual convention of referring to the mode which maximises the form factor as the $\rm TM_{010}$ mode. 

\begin{figure}[htbp]
    \centering
    \includegraphics[width=8.3cm]{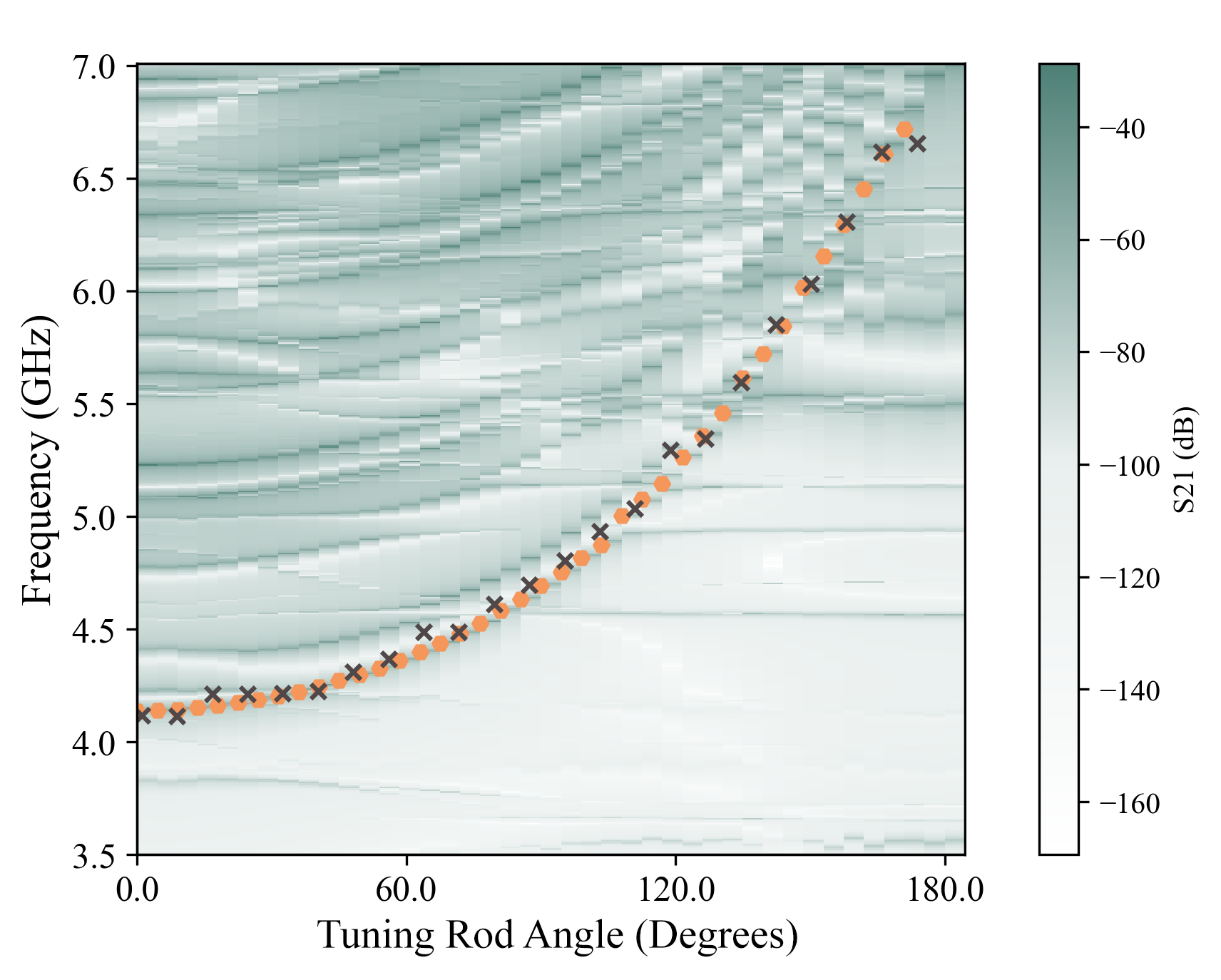}
    \caption{The QSHS cavity mode map, grey grid data taken from ANSYS HFSS simulations, orange circles denote the expected $\rm TM_{010}$ mode frequencies from ANSYS eigenmode simulations and black crosses show the corresponding measured mode frequencies at room temperature with the antennas both overcoupled.}
    \label{fig:ModeMap}
\end{figure}

Cavity simulations herein were conducted using ANSYS electromagnetic finite element analysis simulations, primarily ANSYS HFSS \cite{ansyselectromagnetics}. The QSHS cavity has a tuneable range 4.1 to {\rm 7.5\,GHz}, corresponding to an axion search range approximately $\rm 17$ to $\rm 31\,\mu eV/c^2$. This range was chosen to optimise the operating frequencies to match with devices being developed by the wider QSHS group. The copper cavity mount also has space for three static frequency cavities for further device testing and development. Simulation results and overlaid room temperature measurements of the cavity $\rm TM_{010}$ mode frequency obtained in transmission using a network analyser are shown in Figure \ref{fig:ModeMap}.

\subsection{Cavity and Tuning Mechanism Realisation}
\label{sec:CavityMechanisms}

Three cavities were manufactured, one of aluminium (grade 6061 T6) and two of oxygen-free grade C101 copper. The copper cavities and tuning rod were fabricated by M\&J Engineering \cite{MJEngineering}. The aluminium cavity was fabricated by Eroda \cite{Eroda}. A labelled schematic of the cavity and tuning mechanisms is shown in Figure \ref{fig:CavitySchematic}. One of our two copper cavities was installed as the ADMX sidecar cavity, with the other one installed in QSHS.

The tuning rod is constructed from three pieces: a cylindrical central hollow barrel and two end caps. The end caps are transition-fitted into the cylindrical barrel, resulting in a single hollow cylinder. Small holes in the end caps allow evacuation of the interior. The tuning rod is held in the cavity by sapphire axles \cite{SwissJewels} epoxied into removable copper inserts on each end cap. The primary contact faces of both axles are ceramic deep-groove ball bearings \cite{HalifaxBearings} housed in inserts at the cavity ends. The bottom of the lower axle sits on a ball bearing housed in a 316 stainless support plate bolted to the vertical copper mounting plate.

The field probe antennas both consist of copper RG-405 semi-rigid coaxial cable with the outer conductor and PTFE dielectric stripped off for a few $\rm mm$ at the end. One antenna is kept weakly coupled while the other has adjustable coupling (insertion) to allow near-critical coupling throughout the tuning range. Both antennas are connected to the cavity via friction jounts with beryllium-copper leaf springs housed in  St{\"a}ubli \cite{Staubli} fingerstocks designed by ADMX and inserted into threaded recesses in the top of the cavity. 

The tuning rod position and antenna coupling are both controlled by AttoCube ANR240 piezo motors \cite{AttoCube}. These have encoder resolution {\rm 0.006\,$^{\text{o}}$}; this is expected to correspond to a minimum TM$_{010}$ frequency resolution of {\rm 0.15\,MHz} and should be sufficient to maintain the minimum step size between scans of a third of the current {\rm 3\,dB} bandwidth of the TM$_{010}$ peak. The piezo motors have a maximum dynamic torque at ambient conditions around the axis of {\rm 2\,N\,cm} which is expected to be enough to turn the hollow tuning rod and control the antenna coupling mechanism.

The piezo motors are mounted on the PT2 plate. The large tuning range and hence large change in TM$_{010}$ frequency with tuning rod angle places stringent requirements on the resolution of the tuning mechanism. Both piezo motors are connected to $\approx${\rm 1\,m} halogen-free G10 rods epoxied into couplers that screw directly into the piezo motors. These rods extend downwards from the motors through holes in each of the dilution fridge plates, to a position just above the cavity.

The tuning rod mechanism uses a single-beam clamp-style coupler \cite{HPCGears} to connect the axle to the G10 rod, such that the piezo motor rotation directly correlates to the tuning rod rotation. The rotational-to-linear motion coupler uses a drive nut and corresponding threaded rod. The tube connects to the G10 rod via a triple grub-screw coupler. A slit in the side of this tube allows a screw to be inserted into the threaded rod preventing it rotating and forcing it to instead move vertically as the tube is rotated by the motors. This is such that antenna coupling is possible with no lubrication and such that one full rotation of the piezo motor adjusts the antenna insertion by {\rm 2\,mm}.

\subsection{The Aluminium Cavity}
\label{sec:AlCavity}

An aluminium cavity with geometry identical to that of the copper cavity described in Section \ref{sec:cavity}, manufactured by Eroda\cite{Eroda} was characterised in a test facility at Lancaster University without a tuning rod. These tests led to the designs used for the mechanical clams that secure the QSHS cavity to the support plate and provide a thermal path to cool the cavity. The results are discussed in more detail here \cite{alcavity}.

Aluminium is known to be much easier to machine than copper, and the internal surface of this cavity is visibly better than that of the copper cavity. The aluminium cavity becomes superconducting below the transition temperature $T_c \approx 1.2$\,K, giving a peculiar behaviour of $Q$ and resonance frequency. Our experiments with the Al cavity show that at temperatures not much lower than $T_c$, losses are dominated by quasiparticle excitations and are well described by the BCS theory. The exponential decrease of the quasiparticle number density below $T_c$ results in a 1000-fold increase in the quality factor, as well as a shift in resonance frequency due to the change in the kinetic inductance of the superconductor. 

At very low temperatures, losses due to two-level systems begin to dominate, giving a peak in the quality factor of about
$\rm 2.76\times10^7$ at $\rm 130\,mK$. Unfortunately, this remarkably high Q is unlikely to be utilised in an axion search using aluminium cavities using the Sheffield facility due to the presence of large magnetic fields. Other searches for non-pseudoscalar hidden sector fields could be carried out, and superconducting cavities have also been proposed for use in heterodyne axion searches \cite{PhysRevD.104.L111701}. Furthermore, exploring how aluminium affects the cavity behaviour in both its superconducting and quenched state could provide helpful insights into resonance modes structure and loss mechanisms. Coating cavities in type II high $\rm H_c$ materials is one possibility for significantly increasing the $Q$ factor in strong magnetic fields. 

\section{Receiver Electronics}
\label{sec:receiver} 

\begin{figure}[h!]
	\centering
    \includegraphics[width=8.3cm]{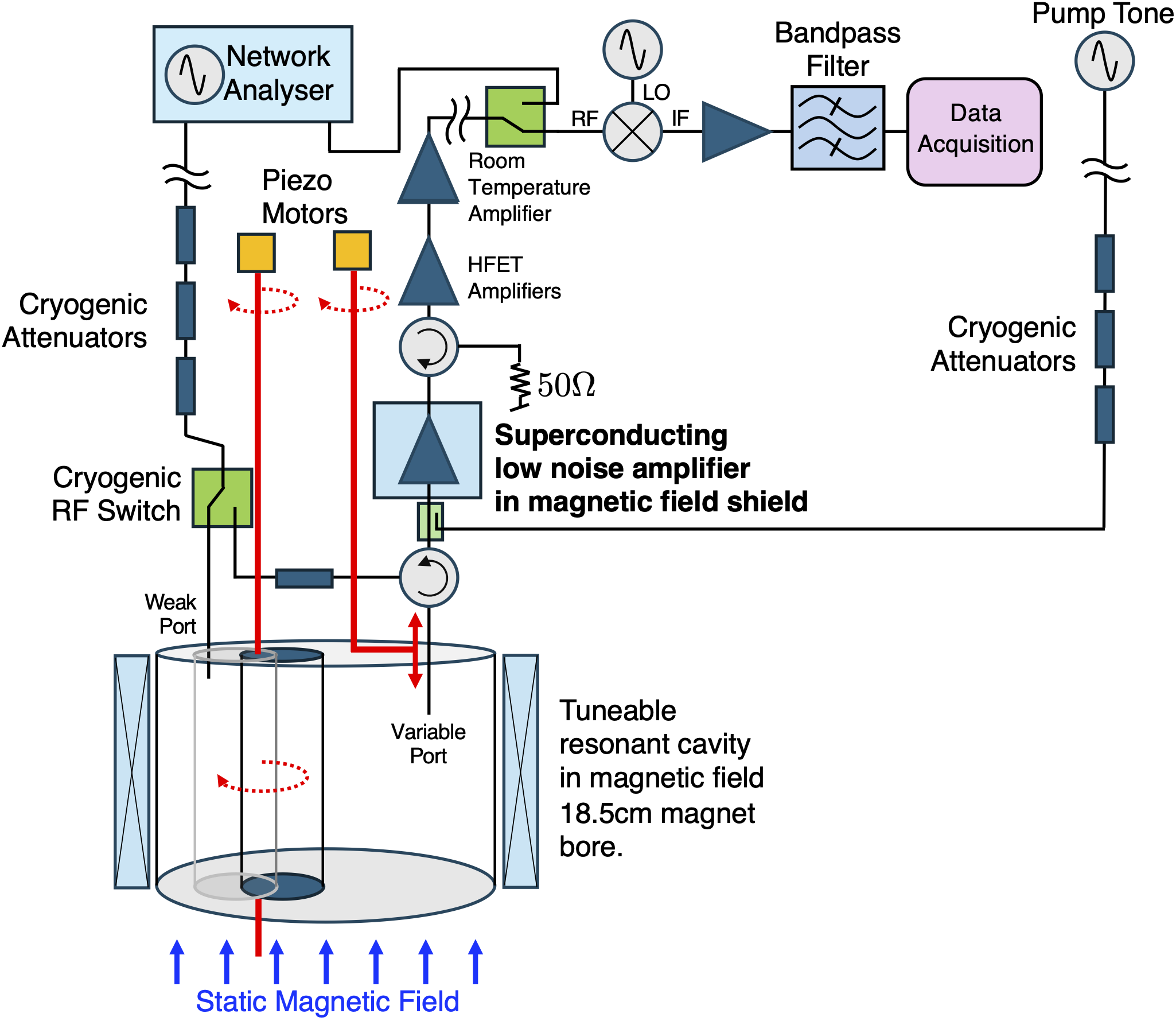}
    \caption{Receiver chain RF output. The demarcation between the DF and the instrument rack is marked with cable breaks. Incoming feeds to the cryostat from external signal sources are attenuated through a string of SMA attenuators built in to SMA feedthroughs that also act as thermal links to the PT1, still plate, and mixing chamber plate stages. The values of the attenuators in these strings are $\rm 20\,dB$ at each of the PT2 and still plate stages and $\rm 10\,dB$ at the mixing chamber plate stage. The pump tone input is used when the ultra-low-noise amplifier is of a parametric type requiring a pump tone.}
    \label{fig:Receiver_Chain}
\end{figure}

Figure \ref{fig:Receiver_Chain} is a schematic of the receiver electronics connected to the resonant cavity system. Vendor web sites supplied in the references contain detailed information of each component in the receiver chain. The variable-coupling port to the cavity feeds an ultra-low-noise amplifier housed in the magnetic field shield via a Low Noise Factory \cite{LowNoiseFactory} LNF-CIC4\_8A circulator. The ultra-low-noise amplifier may be one of several under development in the group, or this stage may be bypassed, in which case the signal feeds directly to the input of the first cryogenic HFET amplifier mounted at the $\rm 4\,K$ stage. A second Low Noise Factory \cite{LowNoiseFactory} LNF-CIC4\_8A circulator provides isolation between the ultra-low-noise amplifier output and the input to the HFET amplifier stages. The HFET amplification consists of two Low Noise Factory LNF-LNC4\_8G \cite{LowNoiseFactory} amplifiers in series, each having a bandwidth of $\rm 4-8\,GHz$ and providing $\rm 40\,dB$ of gain. The output of the second of these HFET amplifiers is fed through a vacuum feed-through to a Low Noise Factory \cite{LowNoiseFactory} LNF-LNR4\_8F room temperature microwave amplification stage, providing a further $\rm 44\,dB$ of gain at room temperature. A cryogenic RF switch allows for swept transmission measurements of the cavity between the weak and variable ports or a reflection measurement off the variable port to verify critical coupling. A room temperature RF switch allows the receiver chain output to feed either the second port of an Anritsu vector network analyser or the room temperature heterodyne receiver apparatus. A Polyphase Microwave \cite{polyphase} IRM4080B image-reject mixer down-converts power from a band around the cavity $\rm TM_{010}$ mode frequency $\nu_0$ to an intermediate frequency (IF) centered at $\rm 10.7\,MHz$. 

The dominant sources of noise in cavity axion searches are Johnson noise from the cavity walls and broadband electronic noise added by the components of the receiver electronics. We therefore aspire to run as cold of a resonator as we can, and to develop and use the lowest possible noise readout electronics. The noise contribution of each component in the receiver chain is divided by the gain of all the previous amplification stages, so that in practice only the first couple of amplification stages make significant contributions to the overall noise temperature.

Calibration of the apparatus consists of conversion of power levels to physical temperatures. Primary calibration will consist of a measurement of the noise temperature of the first Low Noise Factory HFET amplifier. The hot-cold-load method will be used. Having measured the noise temperature of this amplifier, we can switch in an upstream ultra-low-noise device under test (DUT). We employ a technique that has become common in the analysis of receiver electronics. We inject a narrow signal in the centre of the receiver band, through a 50 ohm fixed attenuator to which is afixed a temperature sensor and a heater. We measure the ratio of the power of this peak to the noise floor in the surrounding band as a function of temperature, first with and then without the DUT connected between the attenuator and the first HFET amplifier. The amount by which this ratio increases is the so-called 'signal-to-noise ratio improvement' (SNRI) \cite{ADMX:2025vtp}. The SNRI measurement is then used to infer by how much the first low noise stage is improving the noise temperature of the receiver chain. In reality, it will be a multi-element equivalent circuit whose parameters are constrained by the measured SNRI. Details of the calibration procedure and results are topics for discussion in future publications.

The intermediate frequency (IF) output of the image reject mixer feeds a Mini-Circuits BBP-10.7+ bandpass filter having a $\rm 1\,MHz$ wide pole-free passband. The output of the filter is digitised at a $\rm 125\,MHz$ sampling rate in an AlazarTech ATST146 digitiser. This stand-alone unit was chosen to isolate the DAQ card as far as possible from noise commonly present inside PC instrument towers. The digitised output is fed via a USB-C/Thunderbolt-3 \cite{thunderbolt} connection to a rack mounted PC where analysis of the digitised data takes place. Analysis consists firstly of a digital heterodyne in two orthogonal phases followed by decimation by a power of two, centering the signal band at $\rm 2\,MHz$. This digital signal processing is equivalent to a second heterodyne stage, so our overall electronics consists of a double heterodyne radio receiver circuit.

Averages of the modulus squared of the bins of Fast Fourier transforms of the decimated data form the raw data product. This raw data is stored off-line alongside the results of narrow-band and broad-band sweeps of the cavity in transmission and a header of slow controls data both from low frequency signals associated with the RF electronics readout and other signals read out from the dilution refrigerator and magnet control system.

\subsection{Field Shield}
\label{sec:fieldshield}
	
\begin{figure*}
		\centering
		\includegraphics{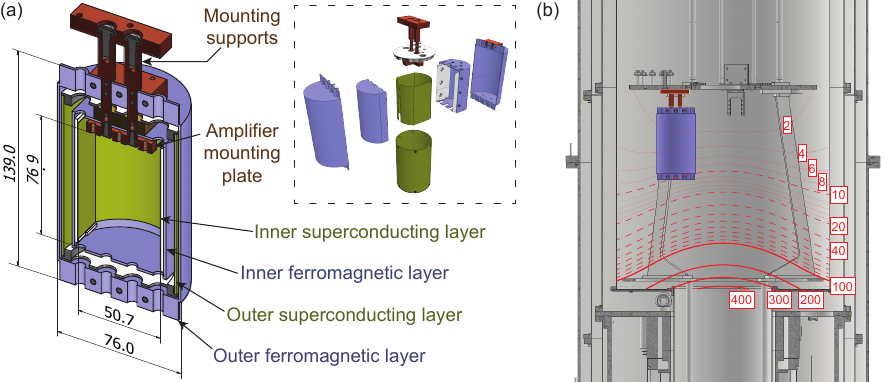}
		\caption{The QSHS field shield. (a) Rendered cross section of the shield, with layers marked. Indicated dimensions are in mm. Inset shows a partly disassembled view. (b) Shield in position beneath the mixing chamber, with superimposed stray field contours (simulated without the shield). Labels indicate field magnitude in mT.}
		\label{fig:FieldShield}
\end{figure*}
	
Superconducting amplifiers must be protected from magnetic fields, which degrade the performance of Josephson junctions and, if they are non-uniform, couple vibrational noise into the electrical signal.
However,  to suppress cable resonances the amplifiers should also be close to the cavity, which is in the high-field region.
To reconcile these requirements, we locate the superconducting electronics inside a passive magnetic field just below the mixing chamber plate.
	
The shield is a multi-layer cylinder (Figure ~\ref{fig:FieldShield}(a)) inspired by that used in the HAYSTAC experiment~\cite{Brubaker2017}.
From the outside, the layers are:
\begin{itemize}
	\item An outer ferromagnetic layer consisting of 1.5\,mm of magnetically soft FeNiMo alloy (Cryophy~\cite{Arpaia2019}), welded from sheet metal and annealed in hydrogen by the manufacturer \cite{magneticshields}.
	\item An outer superconducting layer consisting of a 1.5\,mm seamless Nb tube \cite{specialmetalsfab} with Al end caps \cite{goodfellow}. The Type II superconductor Nb was chosen for the tube because of its high critical field~\cite{Finnemore1966} $B_\mathrm{c1}\approx 199$\,mT. The Type I superconductor Al was chosen for the end caps for ease of machining, despite its lower~\cite{Caplan1965} $B_\mathrm{c}=10.5$\,mT.
	\item An inner ferromagnetic layer, made in the same way as the outer.
	\item An inner superconducting layer consisting of an open cylinder with a longitudinal slit, made from 0.125\,mm Nb foil \cite{goodfellow}.
\end{itemize}
Copper mounting supports anchor a mounting plate inside the shield.
The clear cylindrical volume below this plate has 50\,mm diameter and 76\,mm length.
	
The shield is cooled with the primary magnet unenergized and therefore experiences the Earth's magnetic field $B_\mathrm{earth} \approx 50\,\mu$T.
Most of this field is routed through the outer ferromagnetic layer, ensuring that when the outer superconducting layer is cooled through its transition, the field it encloses is much smaller than $B_\mathrm{earth}$.
To suppress vortex trapping in the superconductor, its cooling path runs through a piece of brass, whose comparatively low thermal conductance should slow down the superconducting transition.
After the cryostat reaches base temperature, the primary coil is ramped up.
By this time, the outer superconducting layer is well below its transition temperature, so its $B_c$ is much larger than the stray field it experiences.
Any flux that does penetrate it is primarily routed through the inner ferromagnetic layer rather than reaching the amplifiers.
The final passive shielding comes from the inner superconducting layer, which does not reduce the field magnitude but does impose a boundary condition that it runs parallel to its surface and therefore predominantly along the cylinder axis.
If this condition is achieved, then the requirement that the $\mathbf{B}$-field have zero divergence and curl means that it will be uniform in this space.
		
Our design is intended to allow fast sample exchange. The ferromagnetic layers are clamshells from which half-shells can be removed without full disassembly (Figure \ref{fig:FieldShield}(a) inset). The inner superconducting layer can be unwrapped and the outer superconducting layer can be slid off. To minimize cable run, there are penetrations at both ends of the shield. In accordance with the principles of good shielding, the penetrations through the ferromagnetic layers are fluted, and the interface between the clamshells is oriented parallel to the stray field lines.
	
We estimate the shielding effectiveness as follows. Since the outer two layers exclude most of the stray field, the field in the amplifier space is mainly the remnant of $B_\mathrm{earth}$ that was trapped inside it during cooldown.
Reference \cite{Mager1970} reports the following estimate, tested in reference \cite{Gubser1979}, for the longitudinal shielding factor of an open double-walled cylinder whose walls have permeability $\mu$:
\begin{align}
	S_\mathrm{L} &\equiv\frac{B_\mathrm{a}}{B_\mathrm{i}} \nonumber \\
				&\approx 4N(L_2/D_2)
	\left( S_1 S_2 \frac{4\Delta \overline{D}^2}{(L_1+\overline{D}/2)D_1^2} 
    +S_1 +S_2 \right) + 1.
\end{align}
Here $B_\mathrm{a(i)}$ is the applied (interior) field, $L_{1(2)}$ and $D_{1(2)}$  the length and diameter of the inner (outer) cylinder, $\Delta$ the cylinder spacing, $\overline{D}$ the average of $D_1$ and $D_2$, $S_i \equiv \frac{\mu d}{D_i}$ the transverse shielding factor of the $i$th shield, $d$ the sheet thickness, and $N(L_2/D_2)$ the demagnetization factor of an ellipsoid with the same aspect ratio as the outer cylinder.
	
Taking dimensions from Figure \ref{fig:FieldShield}(a) and estimating $\mu \approx 10^3$ from the known permeability curve of cryophy~\cite{Arpaia2019} leads to $N(L_2/D_2) \approx 0.22$, $S_1 \approx 23$, $S_2 \approx 20$, and $S_\mathrm{L} \approx 90$. Upon cooling in the Earth's field, the internal field is therefore expected to be
\begin{align}
	B_\mathrm{i} &= \frac{B_\mathrm{earth}}{S_\mathrm{L}} \approx 6 \times 10^{-7}\,\mathrm{T}.
\end{align}
This calculation pessimistically assumes the Earth's field to be vertical and also ignores shielding by the ferromagnetic endcaps.
	
When the primary coil is ramped up, the outer ferromagnetic layer will saturate and therefore lose some shielding effectiveness, but the superconducting layers should remain below their transition fields. Only a small part of the stray field should therefore penetrate the outer superconducting and inner ferromagnetic shields to reach the amplifiers when they are operating. If it turns out that amplifiers under test do not work in the shield, it is an option to obtain a magnetic field probe and measure the field inside the shield as a diagnostic.
			
In addition to shielding effectiveness, another consideration is the magnetic stress on the shield. If we make the pessimistic approximation that the shield is a sphere of infinitely permeability, its magnetic moment in a stray field $\mathbf{B}_\mathrm{stray}$ is~\cite{Jackson1998}
\begin{align}
	\mathbf{m} &= \frac{4\pi}{\mu_0} R^3 \mathbf{B}_\mathrm{stray},
\end{align}
where $R$ is the radius of the notional sphere, taken as half the length of the shield. The force on the shield is thus
\begin{align}
    \mathbf{F} &= \vec{\nabla} (\mathbf{m} \cdot 
        \mathbf{B}_\mathrm{stray}) \nonumber \\
		&= \frac{8\pi}{\mu_0} R^3 B_\mathrm{stray} \vec{\nabla} B_\mathrm{stray}.
\end{align}
At the location of the shield, the stray field of the magnet is $B_\mathrm{stray} \approx 10^{-2}~\mathrm{T}$ with $\left|\vec{\nabla} B_\mathrm{stray} \right| \approx 10^{-1} ~\mathrm{T}~\mathrm{m}^{-1}$ and so, taking $R = 0.07~\mathrm{m}$ from Figure \ref{fig:FieldShield}(a), the magnitude of the force is
\begin{align}
	F	& \approx 7\,\mathrm{N}. 
    \label{eq:F}
\end{align}
	
The same model yields an upper bound on the torque $\tau$.
The magnetic moment of the shield will not be larger than that of a solid sphere, and so
\begin{align}
	\tau 	&= 		\frac{\partial}{\partial{\theta}}(\mathbf{m} \cdot \mathbf{B}_\mathrm{stray}) 	\nonumber \\
	&\lesssim		m B_\mathrm{stray} 	\nonumber \\
	&\lesssim		\frac{4\pi}{\mu_0} R^3 B_\mathrm{stray}^2 \nonumber \\
	&\approx	1.4\,\mathrm{Nm},
\label{eq:tau}
\end{align}
where $\theta$ is the angle between the shield axis and $\mathbf{B}_\mathrm{stray}$. Equations (\ref{eq:F}) and (\ref{eq:tau}) are both upper bounds, and so  $F$ and $\tau$ will be well within the mechanical strength of the shield.

\section{Data Acquisition}
\label{sec:daq} 

With the tuning rod in a fixed position and the variable coupling antenna inserted such that the electronics is critically coupled to the cavity $\rm TM_{010}$ mode, data acquisition consists, first, of wide and narrow band sweeps of the cavity in transmission to determine mode resonant frequency and quality factor. Second, data is acquired through the heterodyne receiver described in Section \ref{sec:receiver}. The raw data consists of Welch estimates of the power spectrum over the bandwidth of the cavity mode. Having acquired sufficient data to achieve a signal to noise ratio, as defined in Equation (\ref{eq:radiometer}), of $\rm SNR=4$, for the signal model under examination, the cavity transmission functions, power spectra, and slow controls data are stored in \texttt{hdf5} format for off-line analysis. The piezo motors are then used to move the tuning rod, and the cycle repeats until the desired range of mode frequencies and axion masses has been probed. A good target is to aim for 90\% data taking live time, opening the apparatus approximately every three months for inspection and diagnostics. If drifts are observed in mode characteristics during long integration times, these times will be broken up into segments and additional measurements of the cavity mode characteristics will be added.

\section{Analysis}
\label{sec:analysis}

We search for excess power in a relatively narrow frequency band within the wider bandwidth of $\rm TM$ modes of the cavity. Frequency-dependent baselines in the power spectra due to the IF bandpass filter and the interactions of the cavity, electronics, circulator impedences, and connecting transmission lines must be subtracted. Peak searches are conducted in the background subtracted data.

Narrow peaks can also be produced by penetration of signals from the ambient electromagnetic environment in the lab through the metal walls of the cryostat and in to the cavity, or along connecting cables. Methods of rejecting non-astrophysical backgrounds include running the receiver with an antenna in the lab free space to see if the signal is present at higher amplitude outside the cavity. 

Anticipated properties of hypothetical dark matter yield other vetos. Axions should convert in the detector with a power proportional to the square of the applied magnetic field, therefore, by varying the strength of the field we can test this anticipated property. Non-axionic hidden sector dark matter may also convert in the detector, but does not require a magnetic field. 

An additional veto for all hidden sector dark matter, not only axions, is to study the precise frequency structure of the peak as a function of time, to see if it exhibits the diurnal or annual modulation that is expected as a consequence of the lab rest frame motion with respect to the dark matter. Such modulation is caused by motion of the lab with dark matter halo as the Earth rotates on its axis and the Sun orbits the galactic centre. This `axion wind' effect should modulate the signal frequency at the $10^{-7}\nu_0$ level.

We may also require that the signal be persistent, although the smoothness of the halo which is required for a persistent signal at constant power is a subject of research, particularly in numerical simulations of dark matter halos. A powerful additional technique is to inform our collaborators at ADMX of a detected signal and run a joint search in spatially-separated detectors.

Finally, astrophysical dark matter signals may contain substructure from components of the halo that have yet to thermalise in the gravitational potential of our galaxy, or have formed bose condensates. Any such substructure would provide further information towards distinguishing astrophysical signals from backgrounds. 

\section{Summary and Next Steps}
\label{sec:summary}

In this paper we have described the initial hardware configuration of the QSHS resonant cavity axion search. The instrument is designed both as a test bed for new low-noise quantum instrumentation and as a search for pseudoscalar hidden-sector dark matter at sensitivities comparable with the expectations for QCD axions. The base temperature of the cavity is $\rm 18.5\,mK$ so that, instrumented with quantum electronics, the ratio of the thermal equilibrium energy, $k_BT$, to the energy of a single quantum of excitation of the resonator, $h\nu$, should be less than unity. 

Next steps include calibration of the instrument sensitivity through measurement of the noise temperature of the receiver. This will involve measuring the noise temperature of the chain starting at the first HFET amplifier. This noise temperature should be dominated by the noise temperature of the first HFET stage. This result can be used to assess the change in noise temperature when additional device prototypes are added before this first HFET amplifier. A low measured noise temperature is an indication that the gain propagated through the receiver chain is sufficient to maintain adequate signal-to-noise ratio at the point where the measured signal is digitised. Such a measurement also provides an absolute calibration of the receiver chain, so that any power excess can be converted to an absolute signal power deposited in the cavity. The apparatus has room for expansion for other experiments and tests of novel experimental techniques. Future experimental configurations will optimise the choice of resonator to optimise signal sensitivity, matched to the receiver electronics. These and other tests of quantum electronics supplied to the QSHS collaboration and searches for axions using the apparatus will be the subject of future collaboration publications. 

\section{Acknowledgements}
\label{sec:acknowledgements}

The QSHS collaboration is funded by the Science and Technology Facilities Council (STFC), Quantum Technologies for Fundamental Physics (QTFP) programme,  grant codes ST/T006811/1 (Sheffield), ST/T006064/1 (NPL), ST/T006099/1 (UCL), ST/T006102/1 (Lancaster), ST/T006145/1 (Liverpool, Oxford),  ST/T006277/1 (Oxford), ST/T006625/1,2 (Oxford, Cambridge) and ST/T006242/1 (RHUL). Further support was provided to the associated `ParaPara' project, funded by STFC, again as part of the QTFP programme, grant codes ST/W006502/1 (Lancaster) and ST/W006464/1 (UCL).  The LLNL group is supported by the U.S. Department of Energy under Contract DE-AC52-07NA27344. LLNL-JRNL-2007312. We wish to thank the Mary and Arthur Hogg Scholarship and Sheffield Alumni Fund for financial support for A. Sundararajan.

\section*{References}
\nocite{*}
\bibliographystyle{ieeetr}
\bibliography{qshs_instrument_paper_2025_08_19.bib}% Produces the bibliography via BibTeX.

\begin{thebibliography}{10}

\bibitem{1933AcHPh...6..110Z}
F.~{Zwicky}, ``{Die Rotverschiebung von extragalaktischen Nebeln},'' {\em Helv.
  Phys. Acta}, vol.~6, pp.~110--127, Jan. 1933.

\bibitem{RevModPhys.90.045002}
G.~Bertone and D.~Hooper, ``History of dark matter,'' {\em Rev. Mod. Phys.},
  vol.~90, p.~045002, Oct 2018.

\bibitem{Yoon_2021}
Y.~Yoon, C.~Park, H.~Chung, and K.~Zhang, ``Rotation curves of galaxies and
  their dependence on morphology and stellar mass,'' {\em Astrophys. J.},
  vol.~922, p.~249, dec 2021.

\bibitem{Wambsganss:1998gg}
J.~Wambsganss, ``{Gravitational lensing in astronomy},'' {\em Living Rev.
  Relativ.}, vol.~1, p.~12, 1998.

\bibitem{ParticleDataGroup:2020ssz}
P.~A. Zyla {\em et~al.}, ``{Review of Particle Physics},'' {\em PTEP},
  vol.~2020, no.~8, p.~083C01, 2020.

\bibitem{Weinberg:1977ma}
S.~Weinberg, ``{A New Light Boson?},'' {\em Phys. Rev. Lett.}, vol.~40,
  pp.~223--226, 1978.

\bibitem{Wilczek:1977pj}
F.~Wilczek, ``{Problem of Strong $P$ and $T$ Invariance in the Presence of
  Instantons},'' {\em Phys. Rev. Lett.}, vol.~40, pp.~279--282, 1978.

\bibitem{Peccei:1977hh}
R.~Peccei and H.~Quinn, ``{CP Conservation in the Presence of Instantons},''
  {\em Phys. Rev. Lett.}, vol.~38, pp.~1440--1443, 1977.

\bibitem{Svrcek:2006yi}
P.~Svrcek and E.~Witten, ``{Axions In String Theory},'' {\em J. High Energy
  Phys.}, vol.~06, p.~051, 2006.

\bibitem{Conlon:2006tq}
J.~P. Conlon, ``{The QCD axion and moduli stabilisation},'' {\em J. High Energy
  Phys.}, vol.~05, p.~078, 2006.

\bibitem{Arvanitaki:2009fg}
A.~Arvanitaki, S.~Dimopoulos, S.~Dubovsky, N.~Kaloper, and J.~March-Russell,
  ``{String Axiverse},'' {\em Phys. Rev. D}, vol.~81, p.~123530, 2010.

\bibitem{Kamionkowski:1992mf}
M.~Kamionkowski and J.~March-Russell, ``{Planck scale physics and the
  Peccei-Quinn mechanism},'' {\em Phys. Lett. B}, vol.~282, pp.~137--141, 1992.

\bibitem{Barr:1992qq}
S.~M. Barr and D.~Seckel, ``{Planck scale corrections to axion models},'' {\em
  Phys. Rev. D}, vol.~46, pp.~539--549, 1992.

\bibitem{Holman:1992us}
R.~Holman, S.~D.~H. Hsu, T.~W. Kephart, E.~W. Kolb, R.~Watkins, and L.~M.
  Widrow, ``{Solutions to the strong CP problem in a world with gravity},''
  {\em Phys. Lett. B}, vol.~282, pp.~132--136, 1992.

\bibitem{Agrawal:2022lsp}
P.~Agrawal, M.~Nee, and M.~Reig, ``{Axion couplings in grand unified
  theories},'' {\em J. High Energy Phys.}, vol.~10, p.~141, 2022.

\bibitem{Agrawal:2024ejr}
P.~Agrawal, M.~Nee, and M.~Reig, ``{Axion couplings in heterotic string
  theory},'' {\em J. High Energy Phys.}, vol.~02, p.~188, 2025.

\bibitem{Abbott:1982af}
L.~F. Abbott and P.~Sikivie, ``{A Cosmological Bound on the Invisible Axion},''
  {\em Phys. Lett. B}, vol.~120, pp.~133--136, 1983.

\bibitem{Preskill:1982cy}
J.~Preskill, M.~Wise, and F.~Wilczek, ``{Cosmology of the Invisible Axion},''
  {\em Phys. Lett. B}, vol.~120, pp.~127--132, 1983.

\bibitem{Dine:1982ah}
M.~Dine and W.~Fischler, ``{The Not So Harmless Axion},'' {\em Phys. Lett. B},
  vol.~120, pp.~137--141, 1983.

\bibitem{Dine:1981rt}
M.~Dine, W.~Fischler, and M.~Srednicki, ``{A Simple Solution to the Strong CP
  Problem with a Harmless Axion},'' {\em Phys. Lett. B}, vol.~104,
  pp.~199--202, 1981.

\bibitem{Zhitnitsky:1980tq}
A.~R. Zhitnitsky, ``{On Possible Suppression of the Axion Hadron Interactions.
  (In Russian)},'' {\em Sov. J. Nucl. Phys.}, vol.~31, p.~260, 1980.

\bibitem{Kim:1979if}
J.~E. Kim, ``{Weak Interaction Singlet and Strong CP Invariance},'' {\em Phys.
  Rev. Lett.}, vol.~43, p.~103, 1979.

\bibitem{Shifman:1979if}
M.~A. Shifman, A.~I. Vainshtein, and V.~I. Zakharov, ``{Can Confinement Ensure
  Natural CP Invariance of Strong Interactions?},'' {\em Nucl. Phys. B},
  vol.~166, pp.~493--506, 1980.

\bibitem{Tobar:2018arx}
M.~E. Tobar, B.~T. McAllister, and M.~Goryachev, ``{Modified Axion
  Electrodynamics as Impressed Electromagnetic Sources Through Oscillating
  Background Polarization and Magnetization},'' {\em Phys. Dark Univ.},
  vol.~26, p.~100339, 2019.

\bibitem{GrillidiCortona:2015jxo}
G.~Grilli~di Cortona, E.~Hardy, J.~Pardo~Vega, and G.~Villadoro, ``{The QCD
  axion, precisely},'' {\em J. High Energy Phys.}, vol.~01, p.~034, 2016.

\bibitem{Sikivie:1983ip}
P.~Sikivie, ``{Experimental Tests of the Invisible Axion},'' {\em Phys. Rev.
  Lett.}, vol.~51, pp.~1415--1417, 1983.
\newblock [Erratum: Phys. Rev. Lett. 52, 695 (1984)].

\bibitem{Sikivie:2020zpn}
P.~Sikivie, ``{Invisible Axion Search Methods},'' {\em Rev. Mod. Phys.},
  vol.~93, no.~1, p.~015004, 2021.

\bibitem{Jackson1998}
J.~D. Jackson, {\em {Classical Electrodynamics}}.
\newblock John Wiley and Sons, Inc., 3~ed., 1998.

\bibitem{oxinstproteox}
``{Oxford Instruments Nanotechnology Tools Ltd., Tubney Woods, Abingdon, OX13
  5QX, U.K.}.'' \url{https://nanoscience.oxinst.com/products/proteoxmx}.
\newblock Accessed: 22\textsuperscript{nd} March 2025.

\bibitem{sumitomo}
``{Sumitomo Cryogenics of Europe, 3 Hamilton Close, Houndmills Industrial
  Estate Basingstoke, Hampshire RG21 6YT, United Kingdom}.''
  \url{https://shicryogenics.com/products/cryocoolers/pulse-tube-cryocoolers/}.
\newblock Accessed: 22\textsuperscript{nd} March 2025.

\bibitem{lakeshore}
``{Lakeshore UK, 1 Mole Business Park, Leatherhead, Surrey KT22 7BA U.K.}.''
  \url{https://www.lakeshore.com/products/categories/overview/temperature-products/ac-resistance-bridges/model-372-ac-resistance-bridge-temperature-controller}.
\newblock Accessed: 22\textsuperscript{nd} March 2025.

\bibitem{ansyselectromagnetics}
ANSYS \textsuperscript{\tiny\textregistered} Electronics HFSS, Release 2023 R1.

\bibitem{MJEngineering}
``{M \& J Engineering Ltd., 39 Finchwell Rd, Handsworth, Sheffield, S13 9AS,
  U.K.}.'' \url{https://www.mandjengineering.co.uk}.
\newblock Accessed: 30\textsuperscript{nd} March 2025.

\bibitem{Eroda}
``{Erodatools Ltd., Unit 4, Laurence Works, Sheffield Road, Penistone,
  Sheffield S36 6HF, U.K.}.'' \url{https://www.erodatoolsltd.co.uk}.
\newblock Accessed: 9\textsuperscript{th} April 2025.

\bibitem{SwissJewels}
``{Swiss Jewel Company, 555 E. City Avenue, 900 Bala Cynwyd, PA 19004,
  U.S.A.}.'' \url{https://www.swissjewel.com/}.
\newblock Accessed: 30\textsuperscript{nd} March 2025.

\bibitem{HalifaxBearings}
``{Halifax Bearings, Unit 9 Margram Business Centre, Horne Street, Halifax,
  West Yorkshire, HX1 5UA, U.K.}.'' \url{https://halifaxbearings.co.uk/}.
\newblock Accessed: 30\textsuperscript{nd} March 2025.

\bibitem{Staubli}
``{St{\"a}ubli Electrical Connectors, 20 Queensbridge, The Lakes, Bedford Road,
  Northampton, NN4 7BF, U.K.}.''
  \url{https://www.staubli.com/global/en/home.html}.
\newblock Accessed: 30\textsuperscript{nd} March 2025.

\bibitem{AttoCube}
``{AttoCube Systems AG, Eglfinger Weg 2, 85540 Haar, Germany}.''
  \url{https://www.attocube.com/en/products/nanopositioners/low-temperature-nanopositioners/anr240reslthv-rotator-360-endless}.
\newblock Accessed: 5\textsuperscript{th} April 2025.

\bibitem{HPCGears}
``{HPC Gears Ltd., Unit 14, Foxwood Industrial Park, Chesterfield, Derbyshire,
  S41 9RN, U.K.}.''
  \url{https://www.hpcgears.com/n/products/couplings/couplings}.
\newblock Accessed: 30\textsuperscript{nd} March 2025.

\bibitem{alcavity}
J.~Esmenda, E.~Laird, I.~Bailey, T.~Gamble, P.~Smith, E.~Daw, and Y.~Pashkin,
  ``Revealing the loss mechanisms of a 3d superconducting microwave cavity for
  use in a dark matter search.'' \url{https://arxiv.org/abs/2503.22637}, April
  2025.

\bibitem{PhysRevD.104.L111701}
A.~Berlin, R.~T. D'Agnolo, S.~A.~R. Ellis, and K.~Zhou, ``Heterodyne broadband
  detection of axion dark matter,'' {\em Phys. Rev. D}, vol.~104, p.~L111701,
  Dec 2021.

\bibitem{LowNoiseFactory}
``{Nellickev\"agen 24412, 63 G\"oteborg, Sweden}.''
  \url{https://lownoisefactory.com}.
\newblock Accessed: 10\textsuperscript{th} April 2025.

\bibitem{polyphase}
``{Technology Service Corporation, 251 18th Street South, Suite 705, Arlington,
  VA 22202, U.S.A.}.''
  \url{https://https://tsc.com/polyphase-microwave-products/}.
\newblock Accessed: 9\textsuperscript{nd} April 2025.

\bibitem{ADMX:2025vtp}
M.~Guzzetti {\em et~al.}, ``{Improved receiver noise calibration for ADMX axion
  search: 4.54 to 5.41{\,}{\,}{\ensuremath{\mu}}eV},'' {\em Phys. Rev. D},
  vol.~111, no.~9, p.~092012, 2025.

\bibitem{thunderbolt}
``Thunderbolt 3 - more speed. more pixels. more possibilities.''
  \url{https://www.thunderbolttechnology.net/sites/default/files/Thunderbolt3\_TechBrief\_FINAL.pdf}.
\newblock Accessed: 15\textsuperscript{th} April 2025.

\bibitem{Brubaker2017}
B.~M. Brubaker, {\em {First results from the HAYSTAC axion search}}.
\newblock PhD thesis, Yale University, 2017.

\bibitem{Arpaia2019}
P.~Arpaia, M.~Buzio, O.~Capatina, K.~Eiler, S.~A. Langeslag, A.~Parrella, and
  N.~J. Templeton, ``{Effects of temperature and mechanical strain on Ni-Fe
  alloy CRYOPHY for magnetic shields},'' {\em J. Magn. Magn. Mater.}, vol.~475,
  no.~August 2018, pp.~514--523, 2019.

\bibitem{magneticshields}
``{Magnetic Shields Ltd., Headcorn Road, Staplehurst TN12 0DS, U.K.}.''
  \url{https://www.magneticshields.co.uk}.
\newblock Accessed: 15\textsuperscript{th} April 2025.

\bibitem{specialmetalsfab}
``{Special Metals Fabrication, 31 Bowlers Croft, Basildon, Enterprise Zone,
  Essex, SS14 3DZ, U.K.}.'' \url{https://special-metals.co.uk}.
\newblock Accessed: 15\textsuperscript{th} April 2025.

\bibitem{goodfellow}
``{Goodfellow Advanced Materials, Ermine Business Park, Huntingdon, Cambs, PE29
  6WR, U.K.}.'' \url{https://www.goodfellow.com}.
\newblock Accessed: 15\textsuperscript{th} April 2025.

\bibitem{Finnemore1966}
D.~K. Finnemore, T.~F. Stromberg, and C.~A. Swenson, ``{Superconducting
  properties of high-purity niobium},'' {\em Phys. Rev.}, vol.~149, no.~1,
  pp.~231--243, 1966.

\bibitem{Caplan1965}
S.~Caplan and G.~Chanin, ``{Critical-field study of superconducting
  aluminum},'' {\em Phys. Rev.}, vol.~138, no.~5A, 1965.

\bibitem{Mager1970}
A.~J. Mager, ``{Magnetic Shields},'' {\em IEEE Trans. Magn.}, vol.~6, no.~1,
  pp.~67--75, 1970.

\bibitem{Gubser1979}
D.~U. Gubser, S.~A. Wolf, and J.~E. Cox, ``{Shielding of longitudinal magnetic
  fields with thin, closely spaced, concentric cylinders of high permeability
  material},'' {\em Rev. Sci. Instrum.}, vol.~50, no.~6, pp.~751--756, 1979.

\bibitem{QSHS:2023jny}
I.~Bailey {\em et~al.}, ``{Searching for wave-like dark matter with QSHS},''
  {\em SciPost Phys. Proc.}, vol.~12, p.~040, 2023.

\bibitem{Graham:2015ouw}
P.~W. Graham, I.~G. Irastorza, S.~K. Lamoreaux, A.~Lindner, and K.~A. van
  Bibber, ``{Experimental Searches for the Axion and Axion-Like Particles},''
  {\em Ann. Rev. Nucl. Part. Sci.}, vol.~65, pp.~485--514, 2015.

\bibitem{Gorghetto:2018myk}
M.~Gorghetto, E.~Hardy, and G.~Villadoro, ``{Axions from Strings: the
  Attractive Solution},'' {\em J. High Energy Phys.}, vol.~07, p.~151, 2018.

\bibitem{Gorghetto:2020qws}
M.~Gorghetto, E.~Hardy, and G.~Villadoro, ``{More axions from strings},'' {\em
  SciPost Phys.}, vol.~10, no.~2, p.~050, 2021.

\bibitem{Gorghetto:2024vnp}
M.~Gorghetto, E.~Hardy, and G.~Villadoro, ``{More axion stars from strings},''
  {\em J. High Energy Phys.}, vol.~08, p.~126, 2024.

\bibitem{Gorghetto:2022sue}
M.~Gorghetto, E.~Hardy, J.~March-Russell, N.~Song, and S.~M. West, ``{Dark
  photon stars: formation and role as dark matter substructure},'' {\em J.
  Cosmol. Astropart. Phys.}, vol.~08, no.~08, p.~018, 2022.

\bibitem{Kolb:1993zz}
E.~W. Kolb and I.~I. Tkachev, ``{Axion miniclusters and Bose stars},'' {\em
  Phys. Rev. Lett.}, vol.~71, pp.~3051--3054, 1993.

\bibitem{Arvanitaki:2019rax}
A.~Arvanitaki, S.~Dimopoulos, M.~Galanis, L.~Lehner, J.~O. Thompson, and
  K.~Van~Tilburg, ``{Large-misalignment mechanism for the formation of compact
  axion structures: Signatures from the QCD axion to fuzzy dark matter},'' {\em
  Phys. Rev. D}, vol.~101, no.~8, p.~083014, 2020.

\bibitem{Agrawal:2023sbp}
P.~Agrawal and A.~Platschorre, ``{The monodromic axion-photon coupling},'' {\em
  J. High Energy Phys.}, vol.~01, p.~169, 2024.

\bibitem{Nelson:2011sf}
A.~E. Nelson and J.~Scholtz, ``{Dark Light, Dark Matter and the Misalignment
  Mechanism},'' {\em Phys. Rev. D}, vol.~84, p.~103501, 2011.

\bibitem{Graham:2018jyp}
P.~W. Graham and A.~Scherlis, ``{Stochastic axion scenario},'' {\em Phys. Rev.
  D}, vol.~98, no.~3, p.~035017, 2018.

\bibitem{Graham:2015rva}
P.~W. Graham, J.~Mardon, and S.~Rajendran, ``{Vector Dark Matter from
  Inflationary Fluctuations},'' {\em Phys. Rev. D}, vol.~93, no.~10, p.~103520,
  2016.

\end{thebibliography}

\end{document}